\documentclass{article}
\usepackage[nonatbib, final]{neurips_2022}
\usepackage[utf8]{inputenc}
\usepackage{amsmath}
\usepackage{float}
\usepackage{algorithm}
\usepackage{algpseudocode}
\usepackage{caption}
\usepackage{enumitem}
\usepackage{graphicx}

\usepackage{hyperref}
\usepackage{caption}
\usepackage{subcaption}
\usepackage[compact]{titlesec}
\titlespacing{\section}{0pt}{2ex}{1ex}    \titlespacing{\subsection}{0pt}{1ex}{0ex}
\titlespacing{\subsubsection}{0pt}{0.5ex}{0ex}
    
\DeclareUnicodeCharacter{2212}{−}

\title{SEIFER: Scalable Edge Inference for Deep Neural Networks}
\author{%
    Arjun Parthasarathy \\
    Crystal Springs Uplands School \\
    Hillsborough, CA 94010 \\
    \texttt{aparthasarathy23@csus.org} \\
    \And
    Bhaskar Krishnamachari \\
    University of Southern California \\
    Los Angeles, CA 90089 \\
    \texttt{bkrishna@usc.edu} \\
}

\begin{document}
\maketitle

\begin{abstract}
    Edge inference is becoming ever prevalent through its applications from retail to wearable technology. Clusters of networked resource-constrained edge devices are becoming common, yet there is no production-ready orchestration system for deploying deep learning models over such edge networks which adopts the robustness and scalability of the cloud. We present SEIFER, a framework utilizing a standalone Kubernetes cluster to partition a given DNN and place these partitions in a distributed manner across an edge network, with the goal of maximizing inference throughput. The system is node fault-tolerant and automatically updates deployments based on updates to the model's version. We provide a preliminary evaluation of a partitioning and placement algorithm that works within this framework, and show that we can improve the inference pipeline throughput by 200\% by utilizing sufficient numbers of resource-constrained nodes. We have implemented SEIFER in open-source software that is publicly available to the research community.
\end{abstract}

\section{Introduction}
\vspace{-0.15in}
Deep Neural Networks (DNNs) have greatly accelerated machine learning across different disciplines, and edge inference is becoming an increasingly popular field with multiple facets \cite{wu2021accelerating}, as sensor-driven computation in many internet of things applications ranging from retail to wearable technology \cite{chen2019deep, biswas2021survey} necessitate DNN inference on the edge. In this paper, we focus on clusters of resource-constrained edge devices that are connected through a wireless network. Such networked \textit{edge clusters} are becoming increasingly common due to their low-cost and scalability at the edge \cite{premkumar2021survey}. 

We address the following questions: How can we take advantage of the characteristics of networked edge clusters to improve the performance of DNN inference while accounting for edge resource constraints? Additionally, can we integrate cloud computing principles such as automated fault-tolerance, high-availability, and container-based abstractions to make edge inference production-viable? 

There are some relevant prior works. DEFER~\cite{parthasarathy2022defer} addresses the partitioning and execution of DNNs on edge clusters, but neither attempts to optimize partitioning and placement to maximize throughput, nor provides support for container orchestration. Couper~\cite{hsu2019couper} does consider containerizing the partitions of a DNN and optimizes the split-points, but doesn't consider the problem of optimizing the placement of the partitions to optimize throughput. Unlike prior edge inference frameworks, we introduce a containerized inference runtime that optimizes both DNN partitioning and placement over a given networked cluster of edge devices.

% Our goal is to increase the practicality of an inference pipeline on a networked edge cluster, with all the benefits of versioning, scalability, and high-availability typically associated with running distributed software in cloud-based data centers. In a networked edge cluster, although we will have a lower inference speed on any individual node, we can take advantage of the parallel computing power enabled by the presence of multiple devices to increase system throughput. To maximize throughput, we have to deploy the model as a partitioned pipeline across multiple devices and identify and minimize the bottleneck latency. While we omit the details of the partitioning and placement algorithm due to space constraints, we show preliminary results indicating that it scales well over a high number of compute nodes to reduce bottleneck latency. 

Our main contribution in this paper is SEIFER, a robust, containerized system to perform inference with a set of model partitions distributed across an edge cluster, integrated into a seamless CI/CD pipeline with Kubernetes. The system is node fault-tolerant and dynamically updates deployments for new model versions. The framework takes into account system resource limitations to provide a lightweight inference runtime. This is all integrated into a plug-and-play standalone cluster solution that provides convenience typically associated with running distributed software in cloud-based data centers. Our code is available at \url{https://github.com/ANRGUSC/SEIFER}. The framework includes a  partitioning and placement algorithm for DNNs across a cluster of edge devices distributed spatially within the same WiFi network, which aims to maximize the throughput of the inference pipeline by minimizing the bottleneck latency. Preliminary results show the throughput increases significantly as we utilize more devices.

\section{System Architecture}
\vspace{-0.15in}
Rather than using the full Kubernetes \cite{kubernetes} framework (which is not suitable for resource-constrained devices), we use microK8s \cite{microK8s}, which is adapted to run on the edge. MicroK8s only modifies the Kubernetes control plane and underlying infrastructure, so we still use Kubernetes constructs for our framework. We create Kubernetes services and take advantage of in-cluster DNS to allow pod communication independent of their lifecycle. In the event of node failure, pods are rescheduled to healthy nodes and the system will continue running. 

\begin{figure}[htbp]
    \centering
    \begin{subfigure}[c]{0.4\linewidth}
        \centering
        \centerline{\includegraphics[scale=0.07]{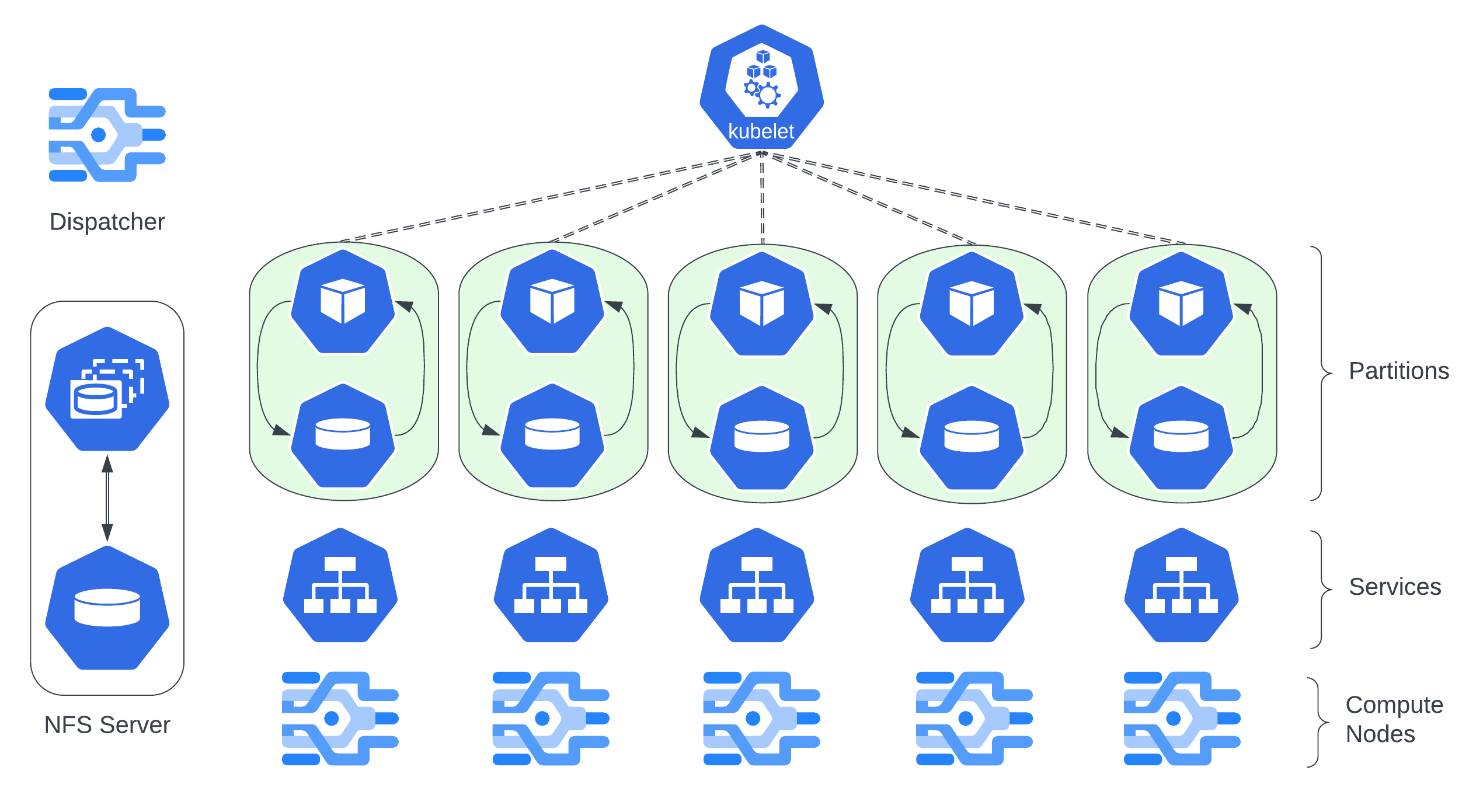}\label{fig:systemoverview}}
        \caption{Kubernetes Cluster Overview}
    \end{subfigure}
    \hfill
    \begin{subfigure}[c]{0.4\linewidth}
        \centering
        \centerline{\includegraphics[scale=0.1]{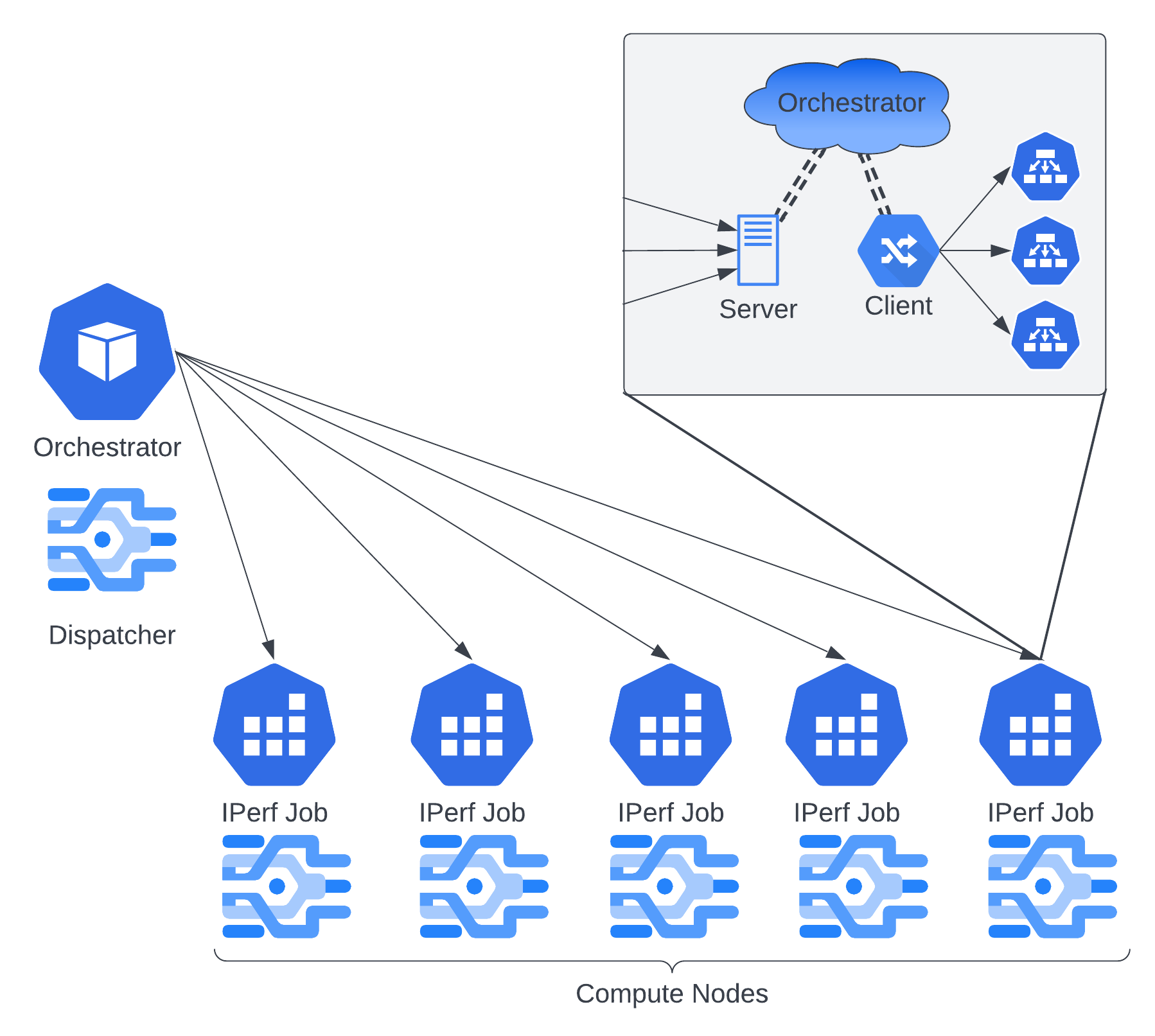}\label{fig:iperfjobs}}
        \caption{IPerf Job Orchestration}
    \end{subfigure}
    \caption{ Container Orchestration in SEIFER }
\end{figure}

\subsection{System Initialization Step}
Upon system startup, the process of leader election starts and the \textit{Dispatcher Node} is chosen. The following events take place:

\begin{enumerate}
    \item \textbf{Scheduling IPerf Jobs}. The system initialization pod launches a job for each node which schedules a pod. Each pod contains a container which runs an IPerf \cite{iperf} server and an IPerf client, which it uses to find the bandwidth between itself and each other node in the cluster. Using a leader-follower architecture, the dispatcher directs each compute node when and where to connect in order to run the IPerf job. Each pod then directs the bandwidth info back to the dispatcher.
    \item \textbf{Dispatcher Pod Deployment}. A deployment is created for the pod running the dispatcher. It is scheduled onto the same node that the system initialization pod is running on, which has been chosen as the leader. The dispatcher pod is configured with the bandwidths between all nodes in the graph.
    \item \textbf{NFS Server}. A cluster-wide NFS server is dynamically provisioned using NFS-Ganesha \cite{nfsganesha}. The NFS server will contain the files necessary to instantiate each node's partition. Since the NFS server has a lifecycle independent of each pod, it will preserve configuration data so that crashed pods can restart their inference runtimes.
\end{enumerate}

\subsection{Configuration Step}
Once the deployment of the dispatcher pod occurs, it runs two init containers which perform the DEFER configuration step as follows:
\begin{enumerate}
    \item \textbf{Partitioning and Placement Container}. The partitioning and placement container first pulls the model from a specified external repository. Using the stored node bandwidth data, the container runs an algorithm we have developed to first partition the model and then to place each partition on a compute node, with the goal of maximizing inference pipeline throughput (which is equivalent to minimizing the bottleneck latency). It quantizes the model and saves the serialized model files to the NFS server. Due to space constraints, we give just a brief overview of the partitioning and placement algorithm below, and present some preliminary results of its empirical performance in the next section.
    \begin{enumerate}
        \item \textbf{Metric:} Latency is given by (amount of data transferred) / (bandwidth).  In order to maximize the throughput of an inference pipeline, we want to minimize the bottleneck latency. Bottleneck latency is calculated as the highest latency between any two nodes in the system. 
        \item \textbf{Partitioning:} Find the model partitions such that the least amount of data is transferred between model layers, and such that each model partition will fit within the compute node's memory.
        \item \textbf{Placement:} Place the partitions such that the ones which transfer the most data are placed on the highest bandwidth edges in the communication graph. Continue this process until all partitions are matched to nodes in the communication graph whose edges have a high enough bandwidth to ensure a low bottleneck latency.
    \end{enumerate}

    \item \textbf{Deploy Container}. The container creates a separate deployment for each inference pod to manage its lifecycle. Each inference pod is scheduled to a certain compute node, and is configured to send its intermediate computed inference to another compute node through a Kubernetes service which hosts an endpoint to the subsequent inference pod.
\end{enumerate}

\subsection{Inference Step}
Each inference pod has two containers:

\begin{enumerate}
    \item \textbf{Inference Runtime Container}. This container instantiates a TFLite \cite{tflite} model from the files on the NFS server. The container contains two FIFOs, which read and write serialized data to the IO container, respectively. Using ZFP \cite{zfp} and LZ4 \cite{lz4} compression, the runtime reads and decompresses data, runs it through the model, and then compresses and writes data. 
    \item \textbf{Inference IO Container}. Contains two FIFOs and two TCP sockets. The FIFOs are used to read and write serialized data to the inference runtime, respectively. One TCP socket acts as a server and receives a connection from the previous compute node, while the other acts as a client and sends the computed inference data to the subsequent compute node. 
\end{enumerate}

\begin{figure}[htbp]
    \centering
     \begin{subfigure}[b]{0.3\textwidth}
        \centering
        \includegraphics[scale=0.04]{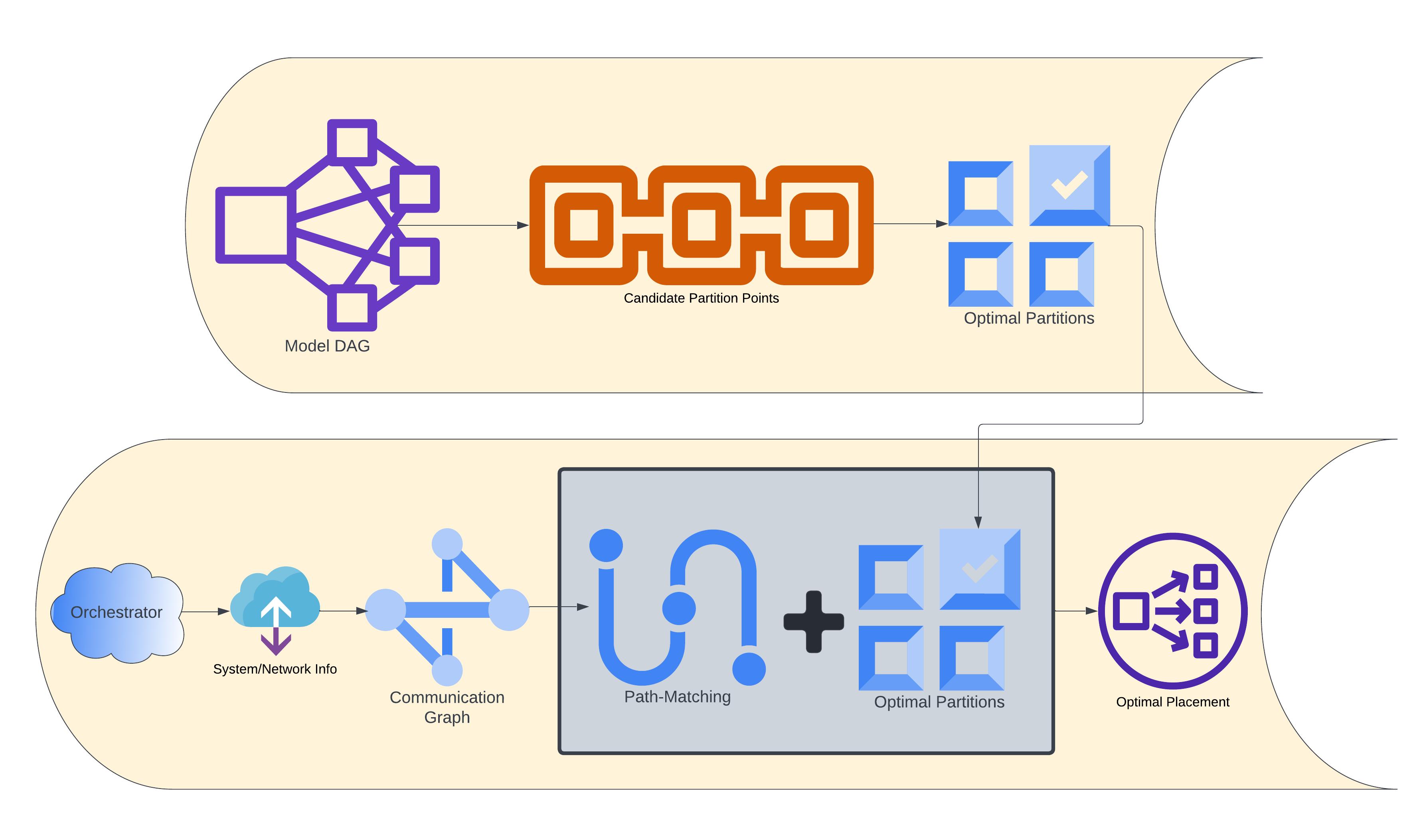}\label{fig:algo}
        \caption{Partitioning and Placement Algorithm}
    \end{subfigure}
    \hfill
    \begin{subfigure}[b]{0.3\textwidth}
        \centering
        \includegraphics[scale=0.1]{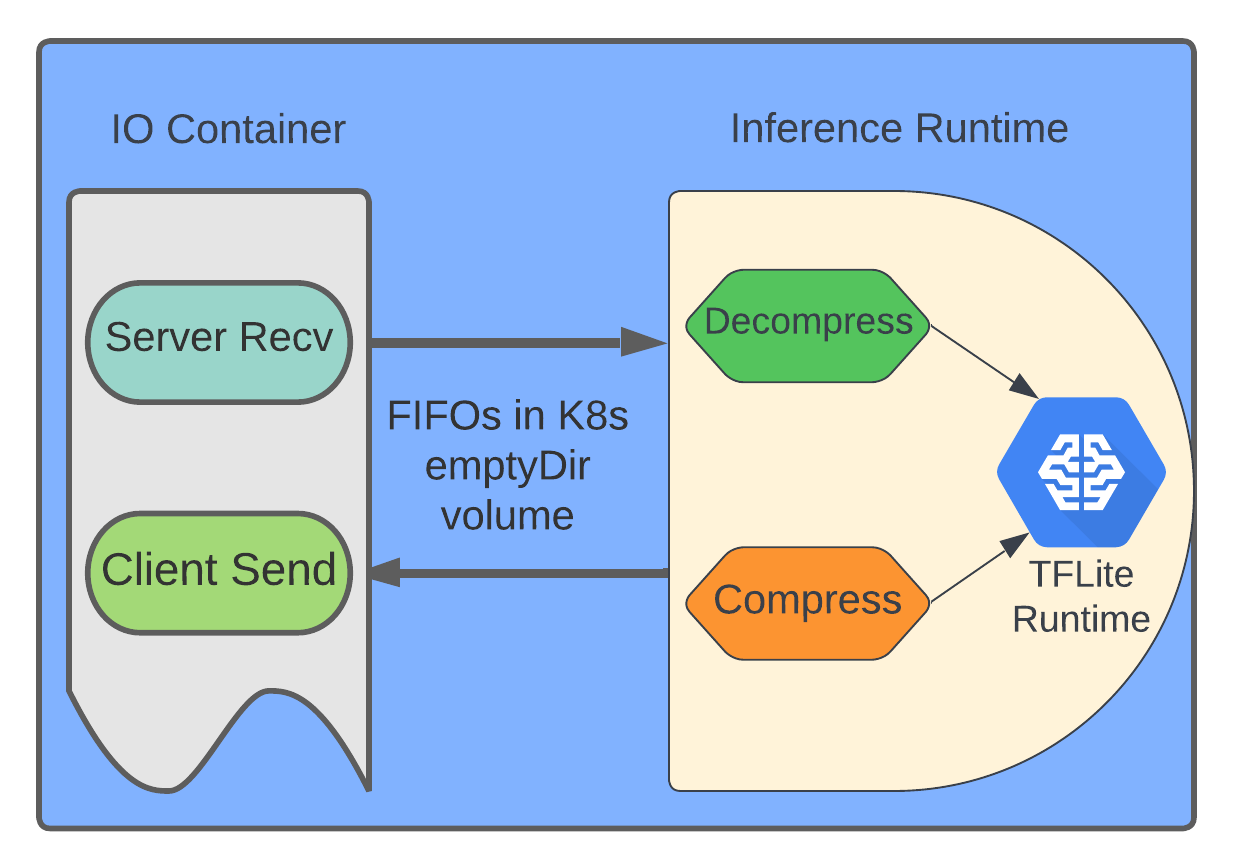}\label{fig:inferencepod}
        \caption{Inference Pod}
    \end{subfigure}
    \hfill
    \begin{subfigure}[b]{0.3\textwidth}
        \centering
        \includegraphics[scale=0.05]{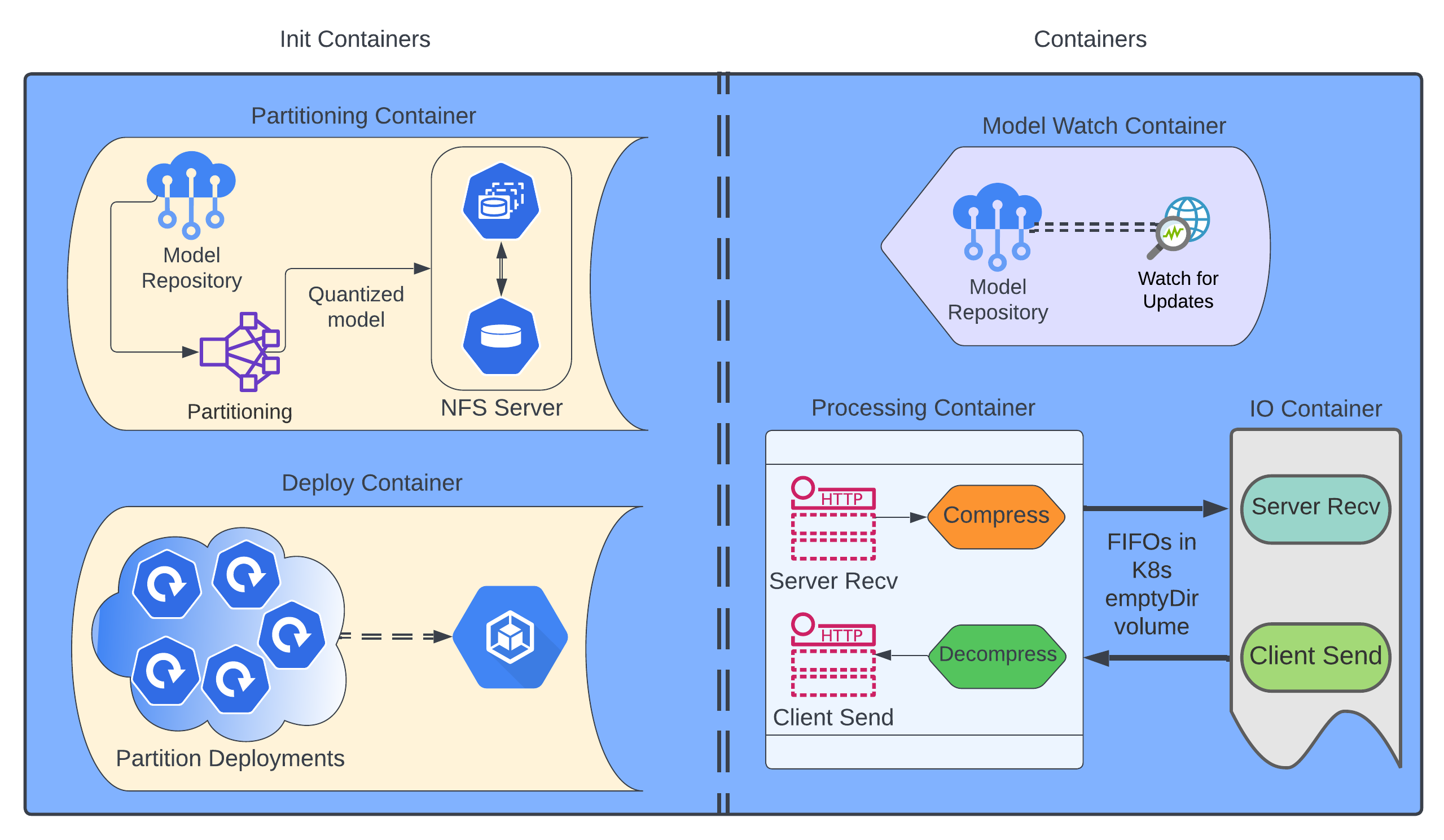}\label{fig:dispatcherpod}
        \caption{Dispatcher Pod}
    \end{subfigure}
    \caption{Key Components of SEIFER}
\end{figure}

Once the inference pods are deployed, the dispatcher runs three containers. In lieu of running on microK8s, these containers can also be packaged and deployed on other edge inference frameworks \cite{akraino, azureiot, awsgreengrass}.

\begin{enumerate}
    \item \textbf{Processing Container}. Runs an HTTP server to read model input. It will convert the model input into the ZFP/LZ4 compressed form used in the system. The container contains two FIFOs, one to send model input data, and the other to receive finished inference results. The container runs an HTTP client which can be configured to send the finished inference data to a certain location.
    \item \textbf{IO Container}. Contains two FIFOs and two TCP sockets. The FIFOs are used to read and write serialized data to the processing container. One TCP socket acts as a server to receive finished inference results from the final compute node, and the other acts as a client to send model input to the first compute node.
    \item \textbf{Model Watch Container}. Watches for updates to the model on the external repository, and if it changes, the container will stop the inference pods and restart model partitioning/placement and deployment. The cluster only needs to be shut down and restarted from the system initialization step if a new node is added.
\end{enumerate}
%\vspace{-0.15in}
\section{Preliminary Results of Partitioning and Placement Algorithm}
\label{section:results}
\begin{figure}[htbp]
\centering
\includegraphics[scale=0.3]{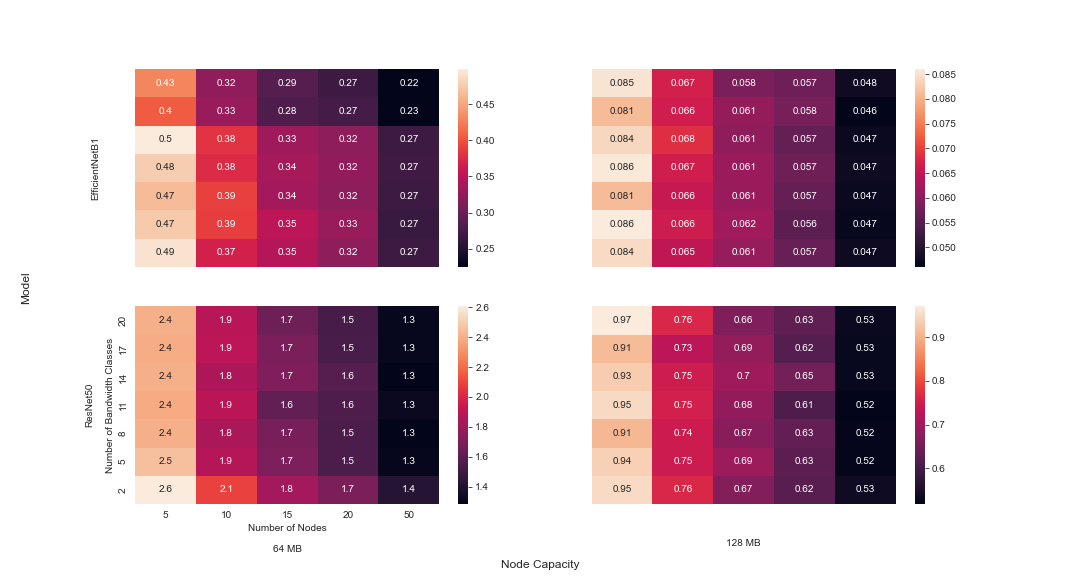}
\caption{Color Map of Bottleneck Latency ($s$) based on Model, Node Capacity, Number of Nodes, and Number of Bandwidth Classes}
\label{fig:bottleneckplotsoptimal}
\end{figure}

\vspace{-0.15in}
To evaluate the partitioning and placement algorithm, we simulated a set of randomly placed edge devices with a wireless network whose link bandwidths are modeled realistically as a function of inter-node distances. The number of ``bandwidth classes" indicates how granular the algorithm is with its bandwidth edge matching. The resulting bottleneck latency from each configuration of model, node capacity, number of nodes, and number of bandwidth classes was run 50 times and averaged.  In Figure~\ref{fig:bottleneckplotsoptimal}, for each model, the lowest bottleneck latency for a given node capacity comes from the combination of the most number of bandwidth classes, number of nodes, and highest node capacity. These results follow from the fact that a larger node and number of nodes allows the partitioning algorithm to have greater choice in selecting the partitions which produce the smallest amount of data transfer. Similarly, a high number of bandwidth classes allows the placement algorithm to
better perform the communication graph edge matching. Across our experiments, we find bottleneck latency improvements of up to 2x, corresponding to a 200\% increase in inference throughput. 
%\vspace{-0.125in}

\vspace{-0.2in}
\section{Ongoing and Future Work}
\vspace{-0.15in}
% We have presented a framework to automate the process of partitioning and placing a model in a distributed manner across a cluster of resource-constrained edge devices, with the goal of maximizing inference throughput. We leverage container orchestration to increase robustness, scalability, and fault-tolerance of the system. We have experimentally shown that we can increase throughput by as much as 200\% using more devices. We have implemented our framework in software and plan to make the source code publicly available for other researchers in the community. 
We are working on the following aspects that we plan to present in future work:
\begin{enumerate}
    \item Present our partitioning/placement algorithm in more detail, and derive characteristics of the DNNs and edge clusters we use that will allow us to model the performance of the algorithm.
    \item Analyze the optimal bottleneck latency and compare the algorithm's performance to an optimal algorithm. We aim to find an approximation ratio for the proposed algorithm.
    \item Compare the algorithm to other strategies such as joint optimization of partition data transfer and communication bandwidth. We will compare the performance of the algorithms with more models and different levels of granularity with the bandwidth matching.
\end{enumerate}

\section{Acknowledgements}
We would like to acknowledge the helpful input and pointers provided by Prof. Anil Vullikanti from the University of Virginia, particularly in directing us to the color-coding $k$-path algorithm that we use to place the partitions across the edge devices. 

\bibliography{main}
\bibliographystyle{acm}
\end{document}